\documentclass[twocolumn,aps,pre,amsmath,amssymb,longbibliography]{revtex4}
\usepackage{graphicx}% Include figure files
\usepackage{dcolumn}% Align table columns on decimal point
\usepackage{bm}% bold math
\usepackage{multirow}
\usepackage{amssymb}
\usepackage{amsmath}
\usepackage{graphicx}
\usepackage{xcolor}
\usepackage[colorlinks,citecolor=blue,linkcolor=red,urlcolor=blue]{hyperref}
\usepackage{indentfirst}
\usepackage{amsmath}
\usepackage{cases}

\begin{document}
\title{Non-equilibrium dynamics of localization phase transition in the non-Hermitian Disordered Aubry-Andr\'{e} model}
\author{Yue-Mei Sun\textsuperscript{1,2}}
\author{Xin-Yu Wang\textsuperscript{1,2}}
\author{Liang-Jun Zhai\textsuperscript{1,2}}

\email{zhailiangjun@jsut.edu.cn}
\affiliation{\textsuperscript{1}The school of mathematics and physics, Jiangsu University of Technology, Changzhou 213001, China}
\affiliation{\textsuperscript{2}The Jiangsu Key Laboratory of Clean Energy Storage and Conversion, Jiangsu University of Technology, Changzhou 213001, China}
\date{\today}
\begin{abstract}
The driven dynamics of localization transitions in a non-Hermitian Disordered Aubry-Andr\'{e} (DAA) model are examined under both open boundary conditions (OBC) and periodic boundary conditions (PBC).
Through an analysis of the static properties of observables, including the localization length ($\xi$), inverse participation ratio ($\rm IPR$), and energy gap ($\Delta E$), we found that the critical exponents examined under PBC are also applicable under OBC.
The Kibble-Zurek scaling (KZS) for the driven dynamics in the non-Hermitian DAA systems is formulated and numerically verified for different local-to-local quench directions.
The hybrid KZS (HKZS) in the overlapping critical region of non-Hermitian DAA and Anderson localization is proposed and numerically confirmed the validity across a local-to-skin quench direction.
%Notably, for the dynamical paths considered, boundary conditions had minimal impact on the applicability of the KZS. However, its evolutionary trajectory will display some variations depending upon whether it traverses the skin-effect region.
This study generalizes the application of the KZS to the dynamical localization transitions within systems featuring dual localization mechanisms.
\end{abstract}
\maketitle

\section{Introduction}
The study of localization phase transition behaviors is a longstanding research topic in condensed matter physics~\cite{Anderson1958,Aubry1980,Biddle070601,Xu2019,Khemani2017,HepengYao2019,Schirmann2024,Ribeiro2024,Jan2023,Alexey2023,Agrawal2020,Agrawalprl2020,Goblot2020,Roy2022,Agrawal2022,
Zhou2023,Smith176601}.
Besides the Anderson model~\cite{Anderson1958}, the localization phase transition has also been found in the quasi-periodic systems,
and various quasi-periodic models have been proposed, such as the Fibonacci model~\cite{Strkalj2021,zhai2021,Guo2024}, the Aubry-Andr\'{e} (AA) model and its various extensions~\cite{Aubry1980,Sarma1988,Biddle2011,Wangprl2020,Liu2021,Gao2024,Ting2022,Longhi20192,Zeng2024,Wang2021,Borgnia1,GYSun2022}.
Studies have shown that there are certain differences in the localization phase transition between Anderson model and quasi-periodic models.
For example, the localization phase transition can even occur for the one-dimensional (1D) AA model~\cite{Aubry1980,Jan2023,Alexey2023}, and the critical exponents for the Anderson and AA models belong to two different universality classes~\cite{Agrawal2020,Wei2019,Wang014206,Cestari2011,Xuan2022}.
These two models have been realized and investigated on various experimental platforms~\cite{Billy2008,Linrj2023,xueprl2022,xueNC2022}, including RLC circuit lattices, acoustic and photonic lattices, single-photon quantum walks, mechanical metamaterials, and cold atoms.

On the other hand, significant advancements in simulating nonequilibrium dynamics in quantum systems have been made~\cite{Wangxh2024,ChenJJ2024,Cheng2399,Xue2024,Zhang2023131,You2024109,Mak2024,Lizhen2024}.
Considerable attention has also been given to the nonequilibrium dynamics in localization transitions~\cite{Yang2017,Heyl2018,Bertini2024,Khan2024}.
Several protocols have been employed to drive a quasi-periodic or disorder system out of equilibrium, such as suddenly changing the Hamiltonian~\cite{Yang2017,Yin2018,Modak2021}, introducing a time-(quasi)periodic driving~\cite{Yang2019,Zhou2021,ZhouDu2021}, or linearly quenching across the localization transition point~\cite{Sinha2019,Liang2024,Xuan2023}.
Sudden quenching of the Anderson and AA models, with initial and post-quench Hamiltonians in different phases, will result in periodic zeros in the Loschmidt echo, indicating the localization-delocalization transition~\cite{Yang2017,Yin2018}.The driven dynamics during a linear quench across the localization transition point can be accurately described by the Kibble-Zurek scaling (KZS)~\cite{Sinha2019,Liang2024,Xuan2023}.

In recent years, the novel phenomena induced by non-Hermiticity have been extensively studied~\cite{Longhi2019,Luo2021,Jing2024,Yuto2021,Lv2022,Jiang2023,Han2023,Sarkar2023,Sunwy2024,XiaoL2024,Daitx2024}, including the non-Hermitian skin effect (NHSE)~\cite{Zhang2109431,Liu2024,Gliozzi2024,YoshidaT2024} and exceptional points~\cite{Wittrock2024,Wang024202,Martinez2018,Aodong706,Wingenbach2024}.
As discussions on non-Hermitian mechanisms have deepened, researchers have integrated these mechanisms into the study of localization~\cite{Jiang2019,Guo2021,Acharya2024,Guocx2024,Kochergin2024}.
The introduction of non-Hermiticity can give rise to a new dimension for tuning localization transitions and uncovers a series of exotic new phenomena~\cite{zhai2021,Jiang2019,zhai2020,Hatano1996,Hatano1998,Luo0153523,Chen144208,Wang024514,Longhi224206,Jazaeri2001,PWang2019,Tang2021,Chaohua123048,YCWang2023,KSuthar2022,Jiang2024,Hamazaki2019}.
For instance, the localization transition in the non-Hermitian quasi-periodic system is accompanied by a transition from real-to-complex eigenvalues, as well as a topological phase transition in the eigenvalue spectrum~\cite{Liu2021,Jiang2019,zhai2020,Hamazaki2019}.
And the NHSE suggests that changes in boundary conditions can result in substantial alterations in the bulk localization properties of non-Hermitian systems~\cite{Guo2021}.
In addition, non-Hermiticity can also lead to critical exponents of the system belonging to different universality classes compared to the Hermitian case~\cite{Luo2021,Zhai2022}.
Experimentally, the introduction of non-Hermiticity can be achieved by using cold atomic traps~\cite{Linrj2023}, specific optical elements\cite{xueprl2022,xueNC2022},  or microwave circuit components~\cite{Jiang2019}. For example, in the single-photon interference network experimental platform, one can simulate different quantum states by encoding the polarization state of photons, and control the selective loss of the polarization state of photons on different paths by setting the angle of the half-wave plate, which reduces the probability amplitude of the system on certain paths and thus induces non-Hermiticity~\cite{xueprl2022,xueNC2022}.

Furthermore, the interplay between non-Hermiticity and non-equilibrium dynamics has also been studied~\cite{Zhou2021,Xuan2023,Zhai2022,Zhaifphy,Cheng2399,Xue2024,Zhang2023131,You2024109,Mak2024,Lizhen2024}.
For example, researchers studied the dynamical evolutions of the non-Hermitian AA model with a complex incommensurate lattice.
They found that Loschmidt echo dynamics cannot detect dynamical phase transitions when the post-quench parameter is in the PT symmetry-broken regime~\cite{Huzh2021}.
Another example is that by applying Floquet time-periodic driving fields to non-Hermitian quasicrystals, one can dynamically control localization transitions and mobility edges as the field parameters vary~\cite{Zhou2021}.
Besides, studies have shown that the KZS remains valid in describing the driven dynamics of the non-Hermitian AA model under both open boundary conditions (OBC)~\cite{Zhaifphy} and periodic boundary conditions (PBC)~\cite{Zhai2022}.

More recently, researchers have increasingly focused on exploring the nonequilibrium dynamics of systems with multiple localization mechanisms~\cite{Liang2024,Xuan2023}.
Bu et al. introduced a Disordered Aubry-Andr\'{e}(DAA) model, combining disorder and quasi-periodic localization, and investigated its driven dynamics~\cite{Xuan2022,Xuan2023}.
Their work proposed a new scaling mechanism, providing a novel perspective on the study of localization phase transition dynamics in systems with coexisting multiple localization mechanisms.
However, the interplay between the non-Hermiticity and the non-equilibrium dynamics in such systems remains unexplored.
Here, we investigate the driven dynamics of localization transitions in a non-Hermitian DAA model with non-reciprocal hopping under both OBC and PBC.
The KZS for the driven dynamics in the non-Hermitian DAA model is constructed and verified numerically across different quench directions.
The rest of the paper is arranged as follows: the non-Hermitian DAA model and phase diagram are introduced in Sec.~\ref{modelphase}. The static scaling properties are studied in Sec.~\ref{cri}.
In Sec.~\ref{KZS}, the KZS for the driven dynamics in the non-Hermitian DAA model under OBC and PBC is constructed and numerically verified.
Then, in Sec.~\ref{HKZS}, the hybrid KZS (HKZS) in the overlapping critical region of non-Hermitian DAA and Anderson localization is proposed and numerically confirmed the validity across a local-to-skin quench direction.
Sec.~\ref{Experiment} introduces an experimental scheme for implementing the non-Hermitian DAA model through time-multiplexed photonic quantum walk system.
A summary is given in Sec.~\ref{sum}.
\section{\label{modelphase}The non-Hermitian DAA Model and phase diagram}
The Hamiltonian of the non-Hermitian DAA model is defined by~\cite{Sun2024}
\begin{eqnarray}
\label{Eq:model}
H &=& -\sum_{j}^{L}{(J_L c_{j}^\dagger c_{j+1}+J_Rc_{j+1}^\dagger c_{j}})+\Delta \sum_{j}^{L}w_j c_j^\dagger c_j\\ \nonumber
&&+(2J_R+\delta)\sum_{j}^{L}\cos{[2\pi(\gamma j+\phi)]c_j^\dagger c_j}.
\end{eqnarray}
Here, $c_j^\dagger (c_j)$ are creation (annihilation) operator of the hard-core boson and $\gamma$ is an irrational number.
$J_L=Je^{-g}$ and $J_R=Je^{g}$ are the asymmetry hopping coefficients.
The parameter $g$ characterizes the extent of asymmetric hopping between lattice sites within the system.
A larger $g$ value corresponds to a greater disparity in asymmetric hopping between the left and right directions.
The adjustment of the value of $g$ can also bring about localized phase transitions and change the degree of localization~\cite{zhai2021,Zhaifphy}.
%In cases where $g$ assumes very large values, non-reciprocal hopping can induce delocalization phenomena in systems initially localized by a disordered potential.
Notably, $g$ plays a crucial role in altering the critical point separating the extended phases from the localization phase~\cite{Jiang2019}. When studying the scaling invariance near the critical point of a non-Hermitian system and determining the relevant critical exponents, it is necessary to first determine the value of $g$ in order to locate the phase transition point. However, previous research has demonstrated that, within a specific range of values, $g$ does not significantly influence the magnitude of the critical exponents~\cite{Zhai2022}. Consequently, in this work, we treat $g$ as a fixed parameter for introducing non-Hermiticity, with a predetermined value of 0.5.
$\Delta$ and $(2J_R+\delta)$ measure the amplitude of disorder and the quasi-periodic potential, respectively.
$w_j\in[-1,1]$ gives the quenched disorder configuration, $\phi\in[0,1)$ is phase of the potential.
We assume $J=1$ as the unity of energy, and $\gamma=(\sqrt{5}-1)/2$.
To satisfy PBC of the quasi-periodic potential, we approximate $\gamma$ as a rational number $F_n/F_{n+1}$ where $F_{n+1}=L$ and $F_{n}$ are the Fibonacci numbers~\cite{Jiang2019,Zhai2022}.
\begin{figure*}[htbp]
\centering
  \includegraphics[width=0.8\linewidth,clip]{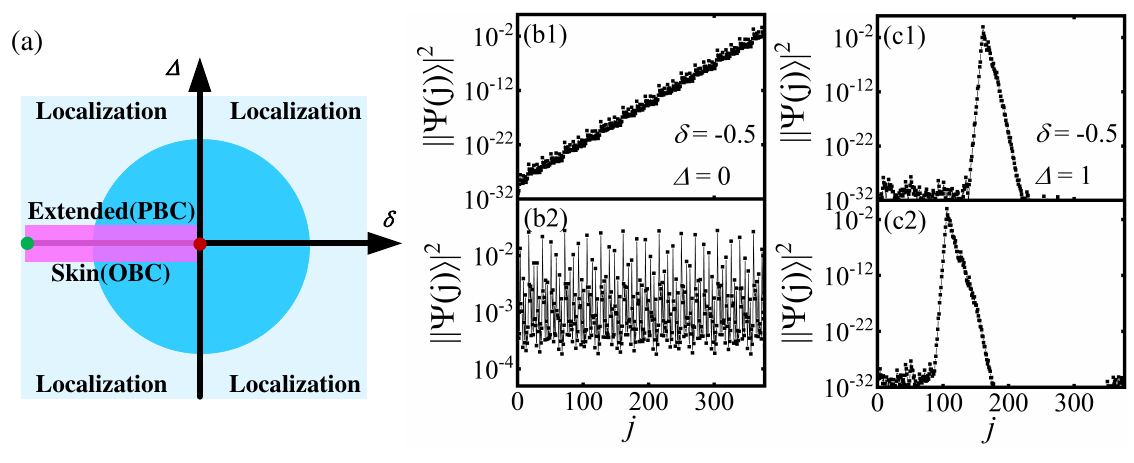}
  \vskip-3mm
  \caption{(a) A schematic representation of the phase diagram for the non-Hermitian DAA model.
  When $\delta=-2J_R$ (denoted by the green point), this model recovers the non-Hermitian Anderson model.
  The light blue region denotes the localization region of non-Hermitian DAA model, the blue region denotes the critical region of non-Hermitian DAA model, the red point denotes the critical point of the non-Hermitian DAA model.
  The purple region denotes the critical region of non-Hermitian Anderson model.
  The part where the blue region and the purple region overlap is the overlapping critical region, in which the critical regions of non-Hermitian DAA and non-Hermitian Anderson localization coexist.
  %The three black arrows represent the three different quenching paths considered in this work.
  When $\delta<0$ and $\Delta=0$, all the eigenstates are extended under PBC, whereas it exhibits the skin-effect phase under OBC. The typical spatial distributions of the state with the lowest real part of the eigenenergy for the non-Hermitian DAA model are shown for $\Delta=0$ under (b1) OBC and (b2) PBC. Similarly, the typical spatial distributions of the state with the lowest real part of the eigenenergy for $\Delta=1$ are illustrated under (c1) OBC and (c2) PBC.
  In these simulations, we use $g=0.5$, $\phi=0$, $\delta=-0.5$, and $L=377$.}
  \label{phase}
\end{figure*}

A schematic phase diagram of the non-Hermitian DAA model near the critical point is illustrated in Fig.~\ref{phase}(a).
%For $\delta>0$, the system will be in a localized state, regardless of the value of $\Delta$.
When $\delta>0$, the system will be in a localized phase regardless of the value of $\Delta$.
%For $\delta<0$, when $\Delta=0$, this model recovers the non-Hermitian AA model~\cite{Jiang2019}. All the eigenstates are extended under PBC~\cite{Zhai2022}, whereas it exhibits the skin-effect phase under OBC~\cite{Jiang2019,Zhaifphy}.
When $\delta<0$ and $\Delta=0$, all the eigenstates are extended under PBC~\cite{Zhai2022}, whereas it exhibits the skin-effect phase under OBC~\cite{Jiang2019,Zhaifphy}.
As illustrated in Figs.~\ref{phase}(b1) and (b2), the spatial distribution of the right eigenvector of the state with the lowest real part of the eigenenergy ($|\Psi(j)\rangle$) with $\delta<0$ and $\Delta=0$ under OBC and PBC are depicted, respectively.
It is shown that the wave function is localized on the boundary under OBC, but it is evenly distributed within a small range under PBC.
When $\delta<0$, any non-zero value of $\Delta$ can induce localization under PBC, whereas a competitive interplay emerges between the skin-effect state and the localized state under OBC.
As shown in Figs.~\ref{phase}(c1) and (c2), when $\Delta=1$ and $\delta<0$, the spatial distributions of states exhibit highly similar localization characteristics under both boundary conditions.

%For $\delta=-2J_R$, this model returns to the non-Hermitian Anderson model~\cite{Hatano1996}.
%However, when $\delta=0$, any finite $\Delta$ can also cause localization at the critical point of the non-Hermitian AA model, but the localization characteristics in this case differ from those of the purely non-Hermitian Anderson model.
The point $(\delta,\Delta)=(0,0)$ represents the critical point of the non-Hermitian DAA model, and the critical region surrounding this point is jointly defined by $\Delta$ and $\delta$.
%In this critical region, the system exhibits localization with a degree that spans from fully extended to fully localized behavior, rapidly varying in response to changes in $\Delta$ and $\delta$.
In this critical region, the system is also in a localized phase, and the correlation length, which characterizes the degree of localization, will exhibit an exponential divergence trend as the parameter $\delta$ or $\Delta$ changes.
Moreover, for $\delta<0$ and infinitesimal $\Delta$, there is a critical region of Anderson localization.
Therefore, for $\delta<0$, the critical region of non-Hermitian Anderson localization and the critical region of non-Hermitian DAA will inevitably overlap with each other, thereby forming an overlapping region.

%For $\delta<0$, any finite $\Delta$ can also cause the system to transition into a localized state.
%As illustrated in Figs.~\ref{phase}(c1) and (c2), when $\Delta=1$ and $\delta<0$, the spatial distribution of localized states exhibits a high degree of similarity under both boundary conditions.

\begin{figure}[tbp]
\centering
  \includegraphics[width=1\linewidth,clip]{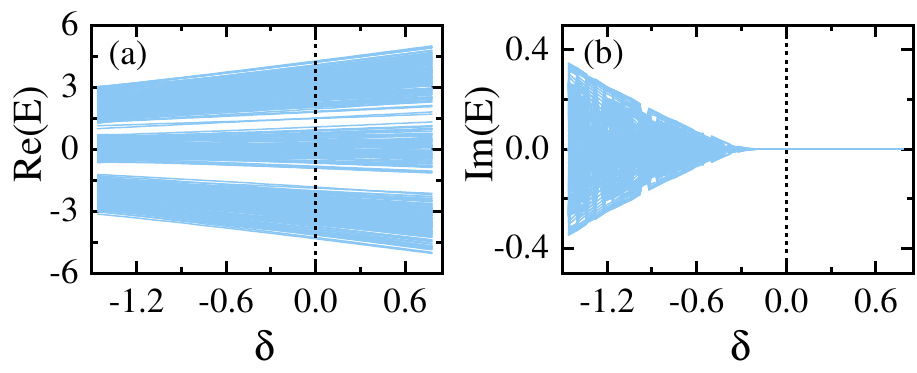}
  \vskip-3mm
  \caption{(a) Real and (b) imaginary parts of energy spectra of the model Eq.(1) under OBC. The black dashed line corresponds to $\delta=0$. Here, we choose $g=0.5$, $\phi=0.2$, $\Delta=0.8$ and $L=377$ in the calculation.
  }
  \label{spectra}
\end{figure}

To discuss the impact of disorder on the localization phase transition in the non-Hermitian AA model, we further explore the energy spectrum of the non-Hermitian DAA model.
The value of $g$ and the magnitude of the potential together lead to the appearance of complex energy spectra in the extended state region of the system's energy spectrum.
Under PBC, the energy spectrum for the non-Hermitian AA model under PBC does not undergo significant changes when disorder is introduced, but it disrupts the correspondence between the transition from real to complex of energy spectrum and the localization transition~\cite{Sun2024}.
Here, Fig.~\ref{spectra} displays the energy spectrum of the non-Hermitian DAA model under OBC.
The results show that, unlike the fully real energy spectrum of the non-Hermitian AA model under OBC~\cite{Jiang2019}, the introduction of disorder leads to the emergence of imaginary parts in the energy spectrum in the region where $\delta<0$.

\section{\label{cri}Static critical properties of the Non-Hermitian DAA model}
In this section, we study the static properties within the critical region, utilizing the localization length ($\xi$), the inverse participation ratio ($\rm IPR$), and the energy gap ($\Delta E$) between the lowest-real-eigenenergy state and the second-lowest-real-eigenenergy state. $\Delta E$ is a key quantity for understanding the stability, dynamics, phase transitions, and potential applications of non-Hermitian systems.

As in the usual quantum criticality, the localization length $\xi$ for the non-Hermitian system is given by~\cite{Sinha2019,Zhai2022}
\begin{equation}
\label{Eq:xiscaling}
   \xi = \sqrt{\sum_{j>j_c}^{L} [( j - j_c )^2 ] P_j},
\end{equation}
where $P_j=||\Psi(j)\rangle|^2$ denotes the probability of the wavefunction at site $j$, and $j_c\equiv\sum jP_j$ represents the localization center.

The IPR is defined as~\cite{Bauer1990,Fyodorov1992}
\begin{equation}
\label{Eq:ipr}
{\rm IPR} = \frac{\sum_{j=1}^L||\Psi(j)\rangle|^4}{\sum_{j=1}^L||\Psi(j)\rangle|^2}.
\end{equation}

In the extended states, the scaling of ${\rm IPR}$ is given by ${\rm IPR}\propto L^{-1}$.
For the localized states or skin effect sates, the scaling is ${\rm IPR}\propto L^0$~\cite{Jiang2019}.

By taking $\Delta$ and $\delta$ as scaling variables simultaneously, the comprehensive finite-size scaling expressions for the three observables are given by
\begin{eqnarray}
\label{Eq:xifull}
  \xi &=& Lf_{1}(\delta L^{\frac{1}{\nu_\delta}},\Delta L^{\frac{1}{\nu_\Delta}}),\\
\label{Eq:IPRfull}
  {\rm IPR} &=& L^{-\frac{s_{\delta}}{\nu_{\delta}}}f_{2}(\delta L^{\frac{1}{\nu_\delta}},\Delta L^{\frac{1}{\nu_\Delta}}),\\
\label{Eq:DEfull}
  {\Delta E} &=& L^{-z_{\delta}}f_{3}(\delta L^{\frac{1}{\nu_\delta}},\Delta L^{\frac{1}{\nu_\Delta}}).
\end{eqnarray}

The function $f_i(.)$ represents the scaling function. These scaling functions, Eqs.~(\ref{Eq:xifull}) to (\ref{Eq:DEfull}), are applicable within the critical region of the non-Hermitian DAA model.
%For the non-Hermitian DAA model, altering either $\delta$ or $\Delta$ will influence the localization state of the system.
Research has found that these two parameters represent two distinct relevant directions, and the critical exponents vary along the $\delta$ or $\Delta$ directions~\cite{Xuan2023,Xuan2022,Sun2024}.
The critical exponents along $\delta$ direction with $\Delta=0$ are $(\nu_\delta,s_\delta,z_\delta)=(1,0.1197,2)$, under both PBC and OBC~\cite{Zhai2022,zhai2021,Zhaifphy}.
In contrast, the scaling exponents along the $\Delta$ direction are $(\nu_\Delta,s_\Delta,z_\Delta)=(0.52,0.0642,2)$ under PBC~\cite{Sun2024}.

%The critical exponents $(\nu_\Delta,s_\Delta,z_\Delta)=(0.52,0.0642,2)$ of the non-Hermitian DAA model along the $\Delta$ direction have been examined under PBC ~\cite{Sun2024}.
\begin{figure}[tbp]
\centering
  \includegraphics[width=1\linewidth,clip]{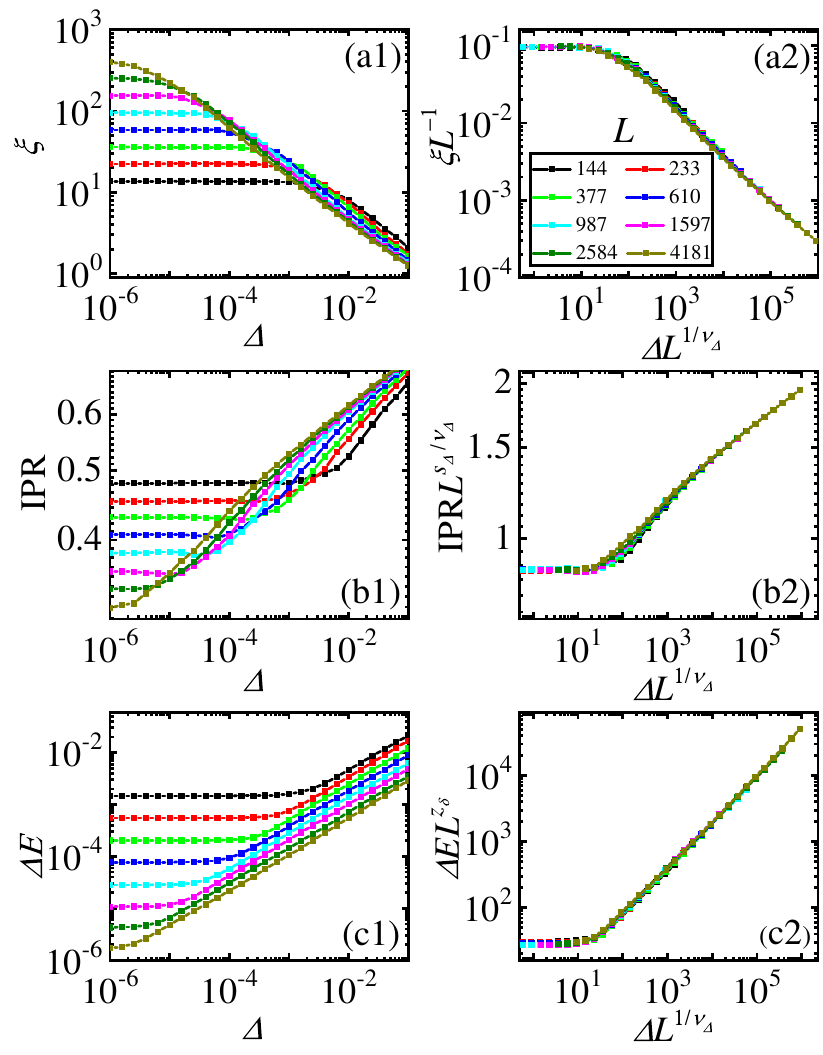}
  \vskip-3mm
  \caption{Scaling properties in the state with the lowest real part of the eigenenergy for fixed $\delta L^{1/\nu_\delta}=-1$ under OBC.
  The curves of $\xi$ versus $\Delta$ before (a1) and after (a2) rescaled for different $L'$s.
  The curves of $\rm IPR$ versus $\Delta$ before (b1) and after (b2) rescaled for different $L'$s.
  The curves of $\Delta E$ versus $\Delta$ before (c1) and after (c2) rescaled for different $L'$s.
   Here, $g=0.5$, and the result is averaged for $1000$ samples. The double-logarithmic scales are used.}
  \label{dLn1}
\end{figure}
%Here, we firstly study the static scaling properties of the non-Hermitian DAA model along the $\Delta$ direction under OBC by employing the exponents obtained from fitting under PBC.
Here, we first verify the applicability of the exponents obtained from fitting along the $\Delta$ direction under PBC to the non-Hermitian DAA model under OBC.
We numerically validate Eqs.~(\ref{Eq:xifull}) to (\ref{Eq:DEfull}) under OBC by maintaining $\delta L^{1/\nu_\delta}$ at a constant value.
Fig.~\ref{dLn1} depict the scaling behavior of $\xi$, $\rm IPR$ and $\Delta E$ as functions of $\Delta$ for $\delta L^{1/\nu_\delta}=-1$.
After rescaling these values according to Eqs.~(\ref{Eq:xifull}) to (\ref{Eq:DEfull}), the resulting curves align closely, thereby confirming the validity of these equations.
%The assignment of the value -1 to $\delta L^{1/\nu_\delta}$ was an arbitrary selection. For the simulation results of another randomly assigned value of -0.5 can be found in Appendix A.
Setting $\delta L^{1/\nu_\delta}$ to -1 is an arbitrary assignment. The simulation results for another randomly selected value of -0.5, which also yielded similar outcomes, are presented in Appendix A.
These results confirm that the same set critical exponent of the non-Hermitian DAA model under PBC is typically sufficient to characterize the critical behavior in the critical region under OBC.

\section{\label{KZS}KZS for the driven dynamics of the non-Hermitian DAA model}
\subsection{General theory of the KZS}
We proceed to investigate the KZS associated with the driven dynamics within the non-Hermitian DAA model.
Our analysis assumes that the system begins in a localized state and is subsequently driven through the critical region by linearly adjusting the distance $\varepsilon$ over time $t$ with a constant rate $R$. The temporal evolution of $\varepsilon$ is described by
\begin{eqnarray}
\label{Eq:varepsilon0}
   \varepsilon = \varepsilon_0-Rt,
\end{eqnarray}
where $\varepsilon$ can represent either $\Delta$ or $\delta$, depending on the specific situation being analyzed, $\varepsilon_0 > 0$ denotes the initial distance from the critical point at $t = 0$.
According to the KZS, if $|\varepsilon| > R^{1/r\nu}$ with the scaling exponent $r = z + 1/\nu$, the system has enough time to adapt to changes in the Hamiltonian, maintaining adiabatic conditions.
Conversely, when $|\varepsilon| < R^{1/r\nu}$, the internal change rate of the system lags behind the rate of the external parameter, indicating the system entry into the impulse region~\cite{Zhong1,Zhong2,Zhong3}.

According to Eq.~(\ref{Eq:model}), when $J_L$, $J_R$, $w_j$ and $\phi$ remain constant, adjusting either variable $\Delta$ or $\delta$ independently can induce a localized phase transition within the system. These two variables represent two independent directions related to the phase transition, and therefore both of these parameters should be incorporated into the full KZS form.
The evolution of the localization length $\xi$ should satisfy~\cite{Xuan2023,Liang2024}
\begin{eqnarray}
\label{Eq:xifullKZS}
  \xi(\Delta, \delta, R)= R^{-\frac{1}{r_\Delta}}f_{4}(\delta R^{-\frac{1}{r_\Delta\nu_\delta}},\Delta R^{-\frac{1}{r_\Delta\nu_\Delta}}),
\end{eqnarray}
where $r_\Delta=z_{\Delta}+1/\nu_{\Delta}$.
Similarly, $\rm IPR$ should satisfy
\begin{eqnarray}
\label{Eq:IPRfullKZS}
  {\rm IPR}(\Delta, \delta, R)= R^{\frac{s_\delta}{r_\Delta\nu_\delta}}f_{5}(\delta R^{-\frac{1}{r_\Delta\nu_\delta}},\Delta R^{-\frac{1}{r_\Delta\nu_\Delta}}).
\end{eqnarray}
Eq.~(\ref{Eq:xifullKZS}) and Eq.~(\ref{Eq:IPRfullKZS}) should be applicable for  a range of $\Delta$ and $\delta$ values in proximity to the critical region of the non-Hermitian DAA model.
Note that for the critical region of the non-Hermitian DAA model, $\nu_{\Delta}$ and $\nu_{\delta}$ are applicable simultaneously, so replacing $r_\Delta$ in Eq.~(\ref{Eq:xifullKZS}) and Eq.~(\ref{Eq:IPRfullKZS}) with $r_\delta=z_{\delta}+1/\nu_{\delta}$, both equations are also valid. In the following content, we will only discuss the situation of $r_\Delta$.

In order to verify the KZS, we can either set some variables to zero or fix the values of certain variables.
When $\Delta=0$, the model~(\ref{Eq:model}) corresponds to the non-Hermitian AA model, and the KZS for the non-Hermitian AA model have been validated~\cite{Zhai2022,Zhaifphy}.
When $\delta = 0$, Eq.~(\ref{Eq:xifullKZS}) and Eq.~(\ref{Eq:IPRfullKZS}) return to
\begin{eqnarray}
\label{Eq:xiKZSdelta0}
  \xi(\Delta, R)&=&R^{-\frac{1}{r_\Delta}}f_{6}(\Delta R^{-\frac{1}{r_\Delta\nu_\Delta}}),\\
\label{Eq:IPRKZSdelta0}
  {\rm IPR}(\Delta, R)&=&R^{-\frac{s_\delta}{r_\Delta\nu_\delta}}f_{7}(\Delta R^{-\frac{1}{r_\Delta\nu_\Delta}}).
\end{eqnarray}
This corresponds to the situation where a dynamic quench of the system is conducted along the direction of $\Delta$, with $\delta$ fixed at 0.
For the case where $\delta\neq0$, we can also fix $\delta R^{-1/r_\Delta\nu_\delta}$ or $\Delta R^{-1/r_\Delta\nu_\Delta}$ to verify Eq.~(\ref{Eq:xifullKZS}) and Eq.~(\ref{Eq:IPRfullKZS}).

\subsection{Numerical results of driven dynamics}
To verify the scaling functions Eq.~(\ref{Eq:xifullKZS}) to Eq.~(\ref{Eq:IPRKZSdelta0}), we numerically solve the schr{\"o}dinger equation for model ~(\ref{Eq:model}) by using the first-order finite difference method in the time direction. The time interval $\Delta t$ is taken as 0.001. And $\varepsilon_0=1.0$ , which is far enough from the critical point at $\varepsilon=0$.
Below, we choose the system size $L = 610$, which is sufficiently large to disregard finite-size effects in realtime simulations.
\begin{figure}[htbp]
\centering
  \includegraphics[width=1\linewidth,clip]{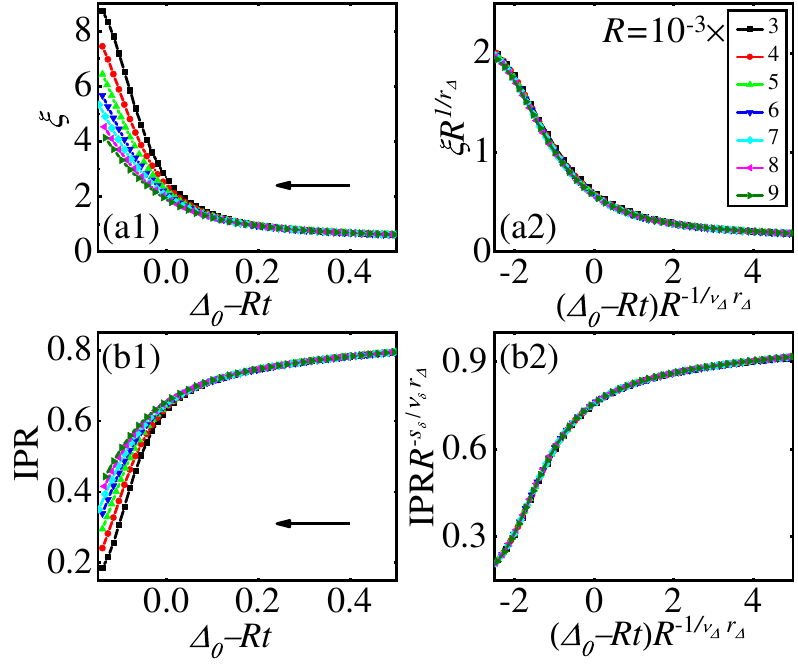}
  \vskip-3mm
  \caption{KZS of driven dynamics in the non-Hermitian DAA model with fixed $\delta = 0$ under OBC.
  Curves of $\xi$ versus $\Delta$ before (a1) and after (a2) rescaling for different driving rates $R$.
  Curves of $\rm IPR$  versus $\Delta$ before (b1) and after (b2) rescaling for different $R$.
  The lattice size is $L=610$, $\phi=0$, $g=0.5$ and one sample of $w_j$ is used.
  $\Delta_0=1.0$.}
  \label{kzsdelta0OBC}
\end{figure}
\begin{figure}[htbp]
\centering
  \includegraphics[width=1\linewidth,clip]{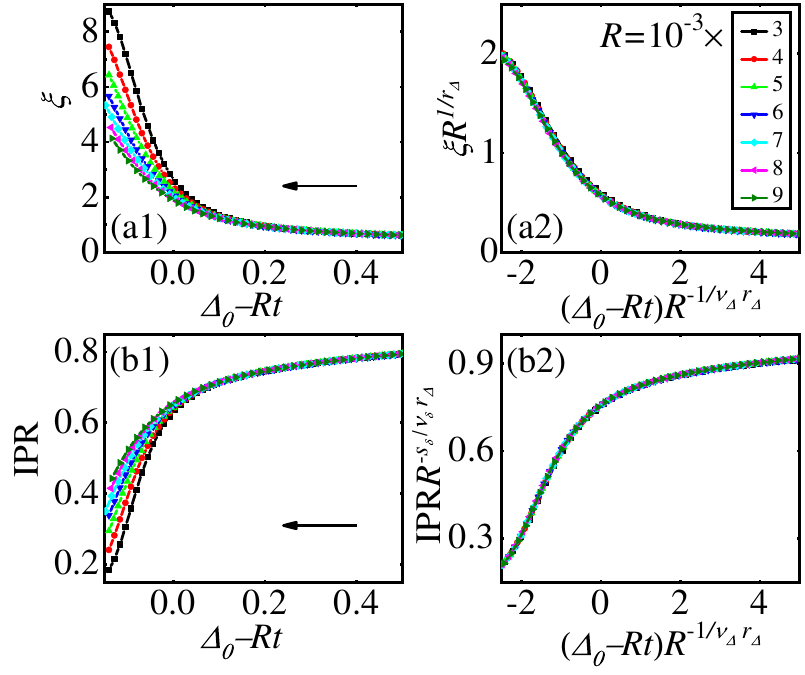}
  \vskip-3mm
  \caption{KZS of driven dynamics in the non-Hermitian DAA model with fixed $\delta = 0$ under PBC.
  Curves of $\xi$ versus $\Delta$ before (a1) and after (a2) rescaling for different driving rates $R$.
  Curves of $\rm IPR$  versus $\Delta$ before (b1) and after (b2) rescaling for different $R$.
  The lattice size is $L=610$, $\phi=0$, $g=0.5$ and one sample of $w_j$ is used.
  $\Delta_0=1.0$.}
  \label{kzsdelta0PBC}
\end{figure}
\begin{figure}[htbp]
\centering
  \includegraphics[width=1\linewidth,clip]{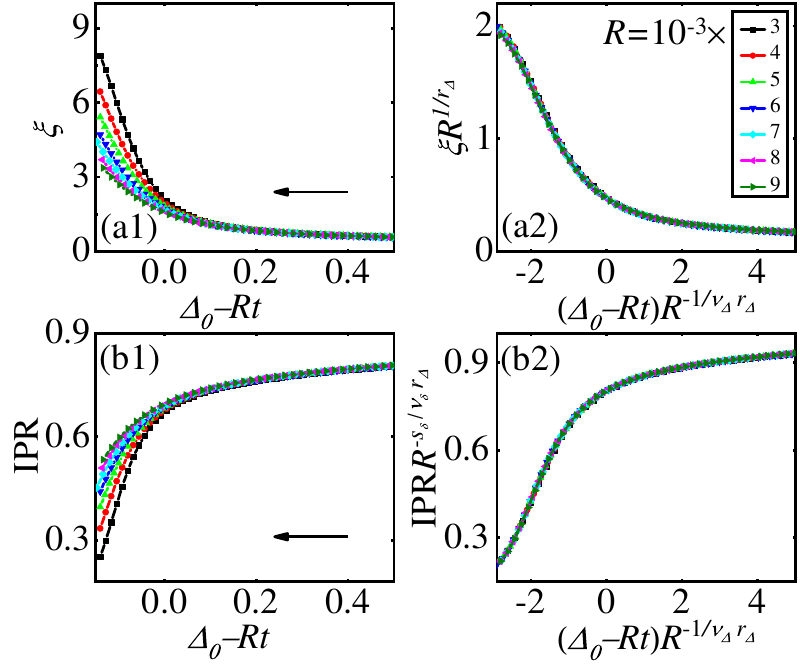}
  \vskip-3mm
  \caption{KZS of driven dynamics in the non-Hermitian DAA model with fixed $\delta R^{-1/r_\Delta\nu_\delta}=0.3$ under OBC.
  Curves of $\xi$ versus $\Delta$ before (a1) and after (a2) rescaling for different driving rates $R$.
  Curves of $\rm IPR$  versus $\Delta$ before (b1) and after (b2) rescaling for different $R$.
  The lattice size is $L=610$, $\phi=0$, $g=0.5$ and one sample of $w_j$ is used.
  $\Delta_0=1.0$.}
  \label{kzsdeltaR03OBC}
\end{figure}
\begin{figure}[htbp]
\centering
  \includegraphics[width=1\linewidth,clip]{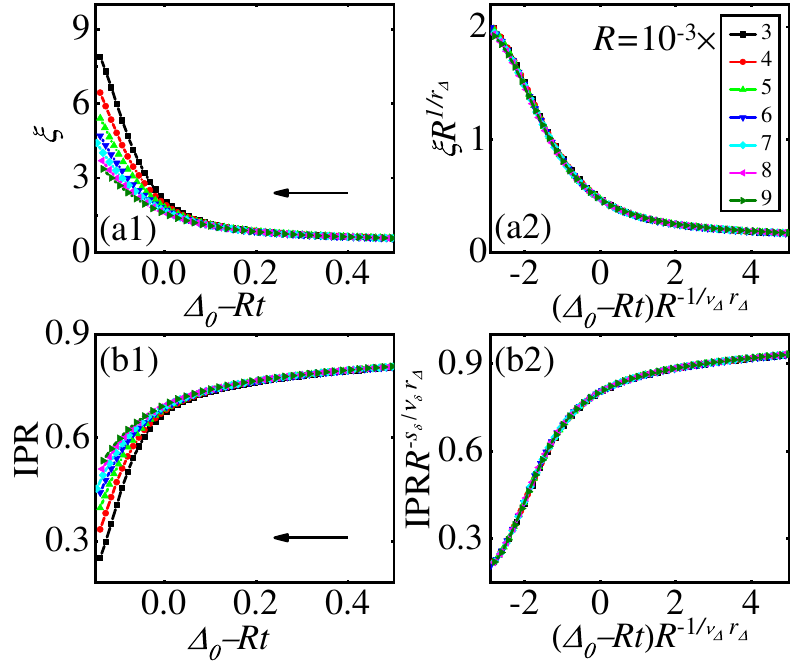}
  \vskip-3mm
  \caption{KZS of driven dynamics in the non-Hermitian DAA model with fixed $\delta R^{-1/r_\Delta\nu_\delta}=0.3$ under PBC.
  Curves of $\xi$ versus $\Delta$ before (a1) and after (a2) rescaling for different driving rates $R$.
  Curves of $\rm IPR$  versus $\Delta$ before (b1) and after (b2) rescaling for different $R$.
  The lattice size is $L=610$, $\phi=0$, $g=0.5$ and one sample of $w_j$ is used.
  $\Delta_0=1.0$.}
  \label{kzsdeltaR03PBC}
\end{figure}

First, we verify the scaling function Eq.~(\ref{Eq:xiKZSdelta0}) and Eq.~(\ref{Eq:IPRKZSdelta0}).
Given that the state with the lowest real part of the eigenenergy often corresponds to the stable state or long-lived state of the system and holds special significance in the process of dynamical evolution, this work will primarily focus on discussing this eigenstate.
We initialize the system in the state with the lowest real part of the eigenenergy at $\Delta_0=\varepsilon_0$, which is deep in the localization state. During the quenching process, the value of the disorder potential $\Delta$ is linearly changed according to Eq.~(\ref{Eq:varepsilon0}). The numerical results for $\xi$ and $\rm IPR$ as functions of $\Delta$ for various driving rates $R$ under OBC are depicted in Fig.~\ref{kzsdelta0OBC}(a1) and Fig.~\ref{kzsdelta0OBC}(b1), respectively.
Observation reveals that the curves corresponding to different $R$ diverge from each other near the critical region, indicating that the system enters the impulse region. However, when these curves are rescaled using the critical exponents of the non-Hermitian DAA model, they converge near the critical region, as illustrated in Fig.~\ref{kzsdelta0OBC}(a2) and Fig.~\ref{kzsdelta0OBC}(b2). This convergence aligns with the predictions of Eq.~(\ref{Eq:xiKZSdelta0}) and Eq.~(\ref{Eq:IPRKZSdelta0}).
Additionally, in Fig.~\ref{kzsdelta0PBC}, the dynamics of $\xi$ and $\rm IPR$ under PBC are plotted. Similarly, the rescaled curves collapse onto each other, confirming Eq.~(\ref{Eq:xiKZSdelta0}) and Eq.~(\ref{Eq:IPRKZSdelta0}).

Then, we verify the KZS forms of $\xi$ and $\rm IPR$ by fixing $\delta R^{-1/r_\Delta\nu_\delta}=0.3$, which is demonstrated in Fig.~\ref{kzsdeltaR03OBC}.
We numerically compute the time evolutions of $\xi$ and $\rm IPR$, initializing the system in the state with the lowest real part of the eigenenergy at $\Delta_0=\varepsilon_0$ for various driving rate $R$ under OBC. The results are presented in Fig.~\ref{kzsdeltaR03OBC}(a1) and Fig.~\ref{kzsdeltaR03OBC}(b1), respectively.
Upon rescaling the quantities according to Eq.~(\ref{Eq:xifullKZS}) and Eq.~(\ref{Eq:IPRfullKZS}), we observe that the curves corresponding to different $R$ converge near the critical region in Fig.~\ref{kzsdeltaR03OBC}(a2) and Fig.~\ref{kzsdeltaR03OBC}(b2).
Similar results for $\xi$ and $\rm IPR$ under PBC are plotted in Fig.~\ref{kzsdeltaR03PBC}.
After rescaling with $R$,  the rescaled curves align with each other as shown in Fig.~\ref{kzsdeltaR03PBC}(a2) and Fig.~\ref{kzsdeltaR03PBC}(b2), confirming Eq.~(\ref{Eq:xifullKZS}) and Eq.~(\ref{Eq:IPRfullKZS}).

\begin{figure}[htbp]
\centering
  \includegraphics[width=1\linewidth,clip]{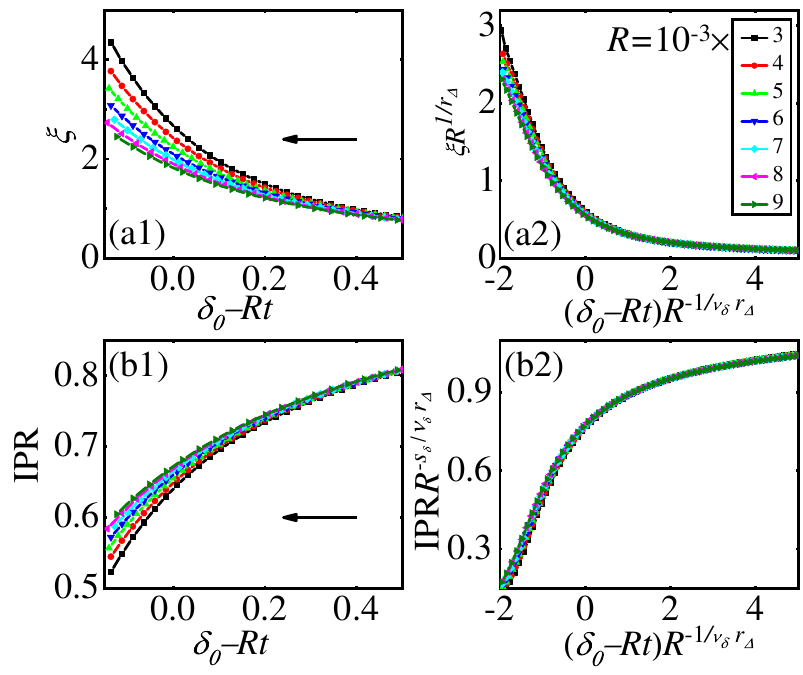}
  \vskip-3mm
  \caption{KZS of driven dynamics in the non-Hermitian DAA model with fixed $\Delta R^{-1/r_\Delta\nu_\Delta}=0.3$ under OBC.
  Curves of $\xi$ versus $\delta$ before (a1) and after (a2) rescaling for different driving rates $R$.
  Curves of $\rm IPR$  versus $\delta$ before (b1) and after (b2) rescaling for different $R$.
  The lattice size is $L=610$, $\phi=0$, $g=0.5$ and one sample of $w_j$ is used.
  $\delta_0=1.0$.}
  \label{kzsDeltaR03OBC}
\end{figure}

\begin{figure}[htbp]
\centering
  \includegraphics[width=1\linewidth,clip]{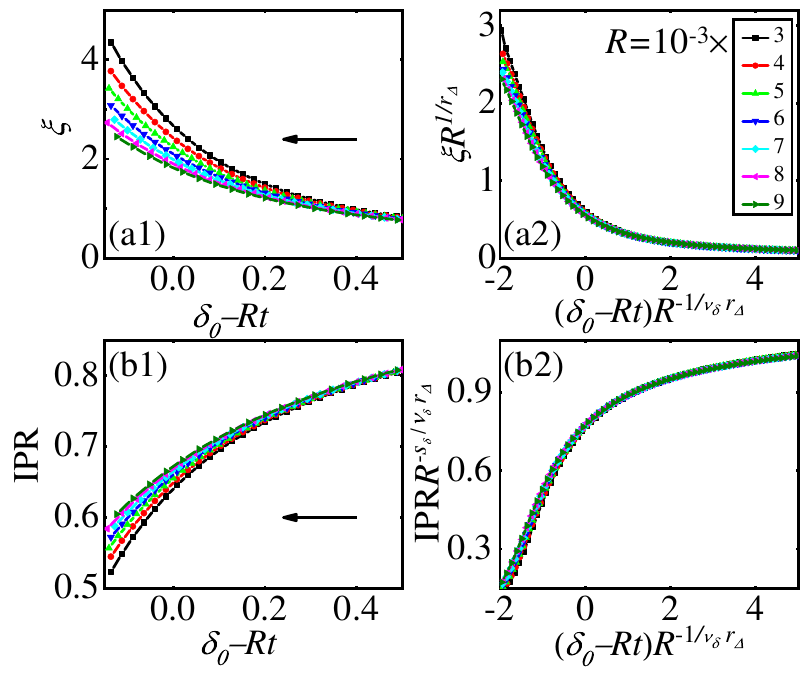}
  \vskip-3mm
  \caption{KZS of driven dynamics in the non-Hermitian DAA model with fixed $\Delta R^{-1/r_\Delta\nu_\Delta}=0.3$ under PBC.
  Curves of $\xi$ versus $\delta$ before (a1) and after (a2) rescaling for different driving rates $R$.
  Curves of $\rm IPR$  versus $\delta$ before (b1) and after (b2) rescaling for different $R$.
  The lattice size is $L=610$, $\phi=0$, $g=0.5$ and one sample of $w_j$ is used.
  $\delta_0=1.0$.}
  \label{kzsDeltaR03PBC}
\end{figure}

We have also numerically verified the KZS for the non-Hermitian DAA model by linearly varying the value of the quasiperiodic potential $\delta$. By fixing $\Delta R^{-1/r_\Delta\nu_\Delta}=0.3$, we calculate the temporal evolution of $\xi$ and $\rm IPR$ for various driving rates $R$, initializing the system in the state characterized by the lowest real part of the eigenenergy at $\delta_0=\varepsilon_0$. Upon rescaling these evolution curves with respect to $R$, we observe that the rescaled curves align with each other, as illustrated in Fig.~\ref{kzsDeltaR03OBC}. This alignment confirms the validity of Eq.~(\ref{Eq:xifullKZS}) and Eq.~(\ref{Eq:IPRfullKZS}).
In Fig.~\ref{kzsDeltaR03PBC}, we present the evolution of $\xi$ and $\rm IPR$ under PBC. After rescaling the curves converge onto one another, in accordance with Eq.~(\ref{Eq:xifullKZS}) and Eq.~(\ref{Eq:IPRfullKZS}).

These findings demonstrate the validity of the KZS in both $\Delta$ and $\delta$ directions near the critical region of the non-Hermitian DAA model.
%Here, the value of 0.3 was randomly chosen. For the simulation results of another randomly assigned value of 0.5, please refer to Appendix B.
Assigning 0.3 to  $\Delta R^{-1/r_\Delta\nu_\Delta}$ is arbitrary. Simulation results for another randomly chosen value (0.5), which showed similar findings, are in Appendix B.

Furthermore, we also observe that for each of the three quenching paths we considered, there is basically no difference in the results of the dynamic numerical simulations between OBC and PBC.
This differs from the conclusions of non-Hermitian AA model studies. In the calculations of the non-Hermitian AA model, although both the driven dynamics under PBC and OBC follow the same KZS, their manifestations exhibit certain differences.
This phenomenon can be understood from the following two aspects. Firstly, all three dynamic evolution processes involve transitions from a localized state, passing through a critical region, and ending up in another localized state.
The localized states are not sensitive to changes in boundary conditions between OBC and PBC.
However, for the non-Hermitian AA model, the phase transition under PBC corresponds to a local-to-extended phase transition, whereas under OBC it manifests as a local-to-skin effect phase transition.
Secondly, for each of the three paths we considered, their energy spectra exhibit similar characteristics under both OBC and PBC. For the case where $\delta=0$ and $\Delta$ is varied, the energy spectra under both OBC and PBC are basically no difference; for the case where $\delta R^{-1/r_\Delta\nu_\delta}$ is fixed and $\Delta$ is varied, the energy spectra under both OBC and PBC are real; for the case where $\Delta R^{-1/r_\Delta\nu_\Delta}$ is fixed and $\delta$ is varied, the energy spectrum undergoes a transition from real to imaginary under both OBC and PBC.
For the non-Hermitian AA model, under PBC, the energy spectrum undergoes a real-to-complex transition, while under OBC, such behavior is absent.


\section{\label{HKZS}HKZS in the overlapping critical region of non-Hermitian DAA and Anderson localization}
As detailed in Sec.~\ref{KZS}, these three quenching paths under consideration do not traverse the skin-effect region.
As previously mentioned, for $\delta<0$, an overlapping critical region emerges where the non-Hermitian DAA and non-Hermitian Anderson localization coexist.
Under PBC, the energy spectra of non-Hermitian DAA systems in the $\delta<0$ region all exhibit imaginary components, making it impossible to analyze driven dynamics.
However, under OBC, since the energy spectra remain real within a certain range, this provides an opportunity to investigate the driven dynamics in the overlapping region.

\subsection{General theory of the HKZS}
In the overlapping critical region where $\delta<0$, scaling functions of non-Hermitian Anderson transitions and the non-Hermitian DAA model are both crucial. To study scaling behavior here, a HKZS is proposed. In a typical case where the overlapping region is assumed to consist of region A and B, it hypothesizes that the driven dynamics in the overlapping region can be described by KZS of both regions simultaneously and a constraint exists between their scaling functions~\cite{Xuan2022,Xuan2023,Yin2017,zhai2018}.
%Overlapping critical regions are common in condensed matter physics, and this law holds general significance, as evidenced by its validation in Hermitian DAA and AA-Stark models, as well as the proposal of hybrid KZS in Yang-Lee edge singularity studies.

Here, we take the critical properties of $\rm IPR$ to illustrate the HKZS.
In the critical region of non-Hermitian Anderson transition, the evolution of the $\rm IPR$ should satisfy
\begin{eqnarray}
\label{Eq:IPRAndersonKZS}
  {\rm IPR}(\Delta, R)= R^{\frac{s_A}{r_A\nu_A}}f_{8}(\Delta R^{-\frac{1}{r_A\nu_A}}).
\end{eqnarray}
where $(\nu_A,s_A,z_A)=(1.99,1.99,2)$ are the critical exponents of the non-Hermitian Anderson transition~\cite{Sun2024}and $r_A=z_{A}+1/\nu_{A}$.

According to the hybrid scaling law, both the scaling functions of $\rm IPR$, i.e., Eqs.~(\ref{Eq:IPRfullKZS}) and (\ref{Eq:IPRAndersonKZS}), are applicable in the critical region where $\delta<0$.
Combining Eqs.~(\ref{Eq:IPRfullKZS}) and (\ref{Eq:IPRAndersonKZS}), the constraint between these two scaling functions should satisfy
\begin{align}
\label{Eq:constraint}
 f_{5}(A,B) = A^{\kappa} f_{9}\left[ B(A)^{\chi} \right]
\end{align}
where $A\equiv{\delta} R^{-1/r_\Delta\nu_\delta}$, $B\equiv{\Delta} R^{-1/r_\Delta\nu_\Delta}$, $\kappa \equiv {r_\Delta \nu_\delta(s_\delta/r_\Delta \nu_\delta- s_A/r_A \nu_A)}$ and $\chi\equiv{r_\Delta}\nu_\delta(1/r_A\nu_A-1/r_\Delta\nu_\Delta)$.
We find that $\chi$ and $\kappa$ include both the critical exponents of non-Hermitian Anderson model and non-Hermitian DAA model, which give the constraint between these scaling functions.
\subsection{Numerical results of driven dynamics}
Here, we numerically verify these scaling tfunctions in the overlapping critical region with $\delta<0$.

By fixing $\delta R^{-1/r_\Delta\nu_\delta}=-0.3$, Eq.~(\ref{Eq:IPRfullKZS}) is firstly verified.
We numerically compute the time evolutions of $\rm IPR$, initializing the system in the state with the lowest real part of the eigenenergy at $\Delta_0=\varepsilon_0$ for various driving rate $R$ under OBC.
The dynamic evolution process involve transitions from a localized state, passing through a skin effect region.
\begin{figure}[htbp]
\centering
  \includegraphics[width=1\linewidth,clip]{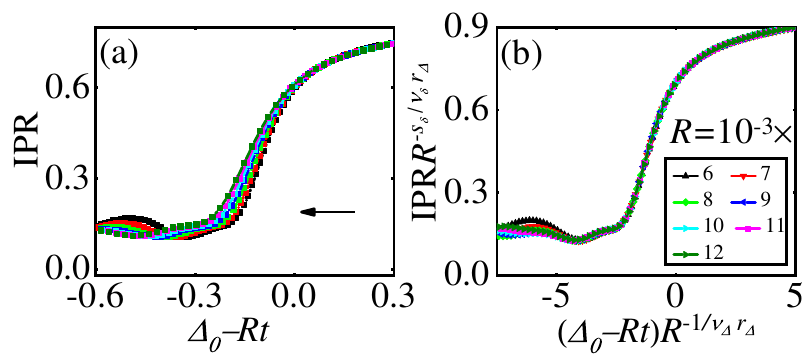}
  \vskip-3mm
  \caption{KZS of driven dynamics in the non-Hermitian DAA model with fixed $\delta R^{-1/r_\Delta\nu_\delta}=-0.3$ under OBC.
  Curves of $\rm IPR$  versus $\Delta$ before (a) and after (b) rescaling for different $R$.
  The lattice size is $L=610$, $\phi=0$, $g=0.5$ and one sample of $w_j$ is used.
  $\Delta_0=1.0$.}
  \label{kzsdeltaRn03OBC}
\end{figure}
\begin{figure}[htbp]
\centering
  \includegraphics[width=1\linewidth,clip]{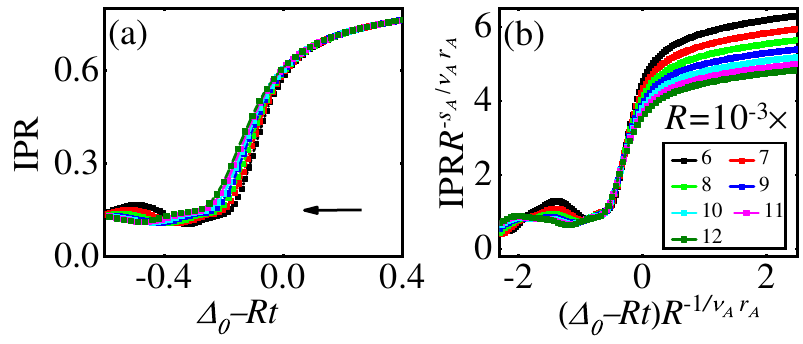}
  \vskip-3mm
  \caption{KZS of driven dynamics in the non-Hermitian DAA model with fixed $\delta=-0.1$ under OBC.
  Curves of $\rm IPR$  versus $\Delta$ before (a) and after (b) rescaling for different $R$.
  The lattice size is $L=610$, $\phi=0$, $g=0.5$ and one sample of $w_j$ is used.
  $\Delta_0=1.0$.}
  \label{kzsdeltan01OBC}
\end{figure}
\begin{figure}[htbp]
\centering
  \includegraphics[width=1\linewidth,clip]{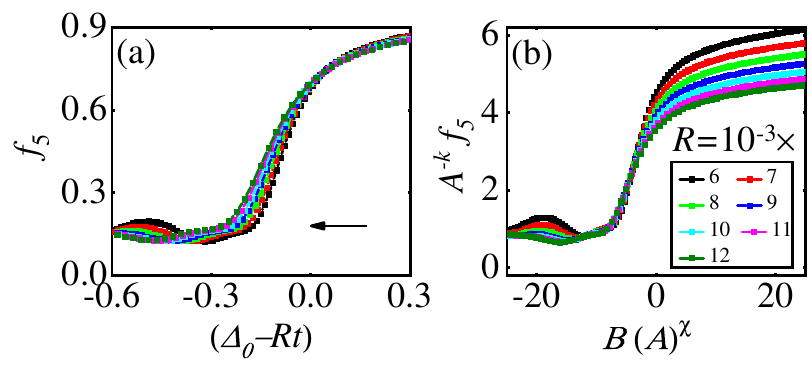}
  \vskip-3mm
  \caption{KZS of driven dynamics in the non-Hermitian DAA model with fixed $\delta=-0.1$ under OBC.
  (a) Curves of $f_{5}$ versus $\Delta$ for different $R$. (b) Curves of  $A^{-\kappa}f_{5}$ versus $B(A)^{\chi}$ for different $R$.
  The lattice size is $L=610$, $\phi=0$, $g=0.5$ and one sample of $w_j$ is used.
  $\Delta_0=1.0$.}
  \label{Hkzsdeltan01OBC}
\end{figure}

In Fig.~\ref{kzsdeltaRn03OBC}, the scaling properties of $\rm IPR$ versus $\Delta$ for $\delta R^{-1/r_\Delta\nu_\delta}=-0.3$ are plotted.
After rescaling according to Eq.~(\ref{Eq:IPRfullKZS}), the rescaled curves collapse onto each other well, confirming Eq.~(\ref{Eq:IPRfullKZS}).
This quenching path crossing the skin-effect region reveals that the skin effect creates depression zones in the IPR curves, which suggests the emergence of a skin-effect state~\cite{Zhaifphy}.
It exhibits certain differences from the evolutionary results of the three previously discussed quenching paths that transition from localized states to localized states.
Therefore, the choice of boundary conditions does not significantly affect the applicability of the KZS, however, during the evolution process of the system, its evolutionary trend will exhibit some differences depending on whether it passes through the skin-effect region.

Next, we take a fixed $\delta<0$ and vary $R$ to verify Eq.~(\ref{Eq:IPRAndersonKZS}).
The numerical results for $\delta=-0.1$ are plotted in Fig.~\ref{kzsdeltan01OBC}.
The collapse of the rescaled curves shown in Fig.~\ref{kzsdeltan01OBC} (b) also confirms Eq.~(\ref{Eq:IPRAndersonKZS}).
Therefore, numerical results in Fig.~\ref{kzsdeltaRn03OBC} as well as Fig.~\ref{kzsdeltan01OBC} confirm the first hypothesis of the hybrid scaling law.

Finally, we also verify Eq.~(\ref{Eq:constraint}) using a fixed $\delta<0$ with varying $R$.
The numerical results of $f_5=\rm IPR R^{-s_\delta/r_\Delta\nu_\delta}$ as a function of $\Delta$ for various $R$ are plotted in Fig.~\ref{Hkzsdeltan01OBC} (a).
By rescaling $\Delta$ as $B(A)^{\chi}$, we find that the rescaled curves collapse very well, verifying Eq.~(\ref{Eq:constraint}) and the second hypothesis of the hybrid scaling law.

\section{\label{Experiment}Experimental scheme}
%Recently, the validity of the KZS in non-Hermitian systems has been experimentally verified through single-photon interference. Our research results are also expected to be verified in similar platforms.
Recently, the validity of the KZS in non-Hermitian systems has been experimentally verified through a time-multiplexed photonic quantum walk system ~\cite{xuekzs2021}. Building on this progress, our research results are also expected to be verified in similar platforms.
This approach employs time-multiplexed encoding, where photon polarization and arrival time encode spatial and internal degrees of freedom. Non-Hermiticity is introduced via polarization-dependent loss, dynamically controlled by an electro-optic modulator (EOM).
Our DAA model is constructed by incorporating quasi-periodic potential through position-dependent phase operators~\cite{xueprl2022} and by introducing disorder via EOM-driven random modulation of coin operation angles~\cite{xueNC2022}. The quantum walk dynamics are governed by the Floquet operator~\cite{xueprl2022,xueNC2022}. The photon state is initialized and evolved through the quantum walk, with system parameters gradually varied to traverse the phase transition point. Photon distributions are recorded using avalanche photodiodes (APDs) to measure dynamic observables. Defect density and fluctuations are analyzed to validate KZS behavior, with critical exponents $\nu$ and $z$ extracted from power-law fits to confirm the universality class of the phase transition. This framework provides a platform for investigating the interplay of disorder and quasi-periodicity in non-equilibrium quantum systems, deepening insights into localization phenomena.

\section{\label{sum}summary}
In summary, we have conducted a study of the static scaling behavior and the driven dynamics of localization transitions in the non-Hermitian DAA model, considering both OBC and PBC.
By examining the static behavior of $\xi$, $\rm IPR$ and $\Delta E$, we have demonstrated that the same critical exponents observed under PBC are also valid under OBC.
We constructed the KZS for driven dynamics of the non-Hermitian DAA systems and numerically confirmed its validity across three different local-to-local quench directions. The HKZS in the overlapping critical region of non-Hermitian DAA and Anderson localization has been proposed and numerically confirmed the validity across a local-to-skin quench direction.
For the dynamical paths we considered, the boundary conditions do not have a significant impact on the applicability of the KZS. However, during the evolution process of the system, its evolutionary trend will exhibit some differences depending on whether it passes through the skin-effect region.
Our work generalizes the application of the KZS to the dynamical localization transitions within systems featuring dual localization mechanisms.

On-site gain and loss constitute another crucial aspect of non-Hermiticity~\cite{Tomasi2023,Longhi2023}, with numerous studies highlighting its distinct influence on localization compared to nonreciprocal effects. Hence, investigating the non-Hermitian DAA model that incorporates on-site gain and loss stands as a potential and fruitful extension of this paper.
Additionally, the absence of a mobility edge in the excited states of DAA model has been demonstrated~\cite{Xuan2022,Sun2024}. However, the introduction of dynamic driving to systems that do possess a mobility edge may result in unique phenomena, necessitating further exploration and discussion.
\section*{ACKNOWLEDGMENTS}
This work is supported by the National Natural Science Foundation of China (Grant No. 12274184), the Qing Lan Project, the National Natural Science Foundation of China (Grant No. 12404105) and the Natural Science Foundation of the Jiangsu Higher Education Institutions of China (Grant No. 24KJB140008).
\begin{figure}[tbp]
\centering
  \includegraphics[width=1\linewidth,clip]{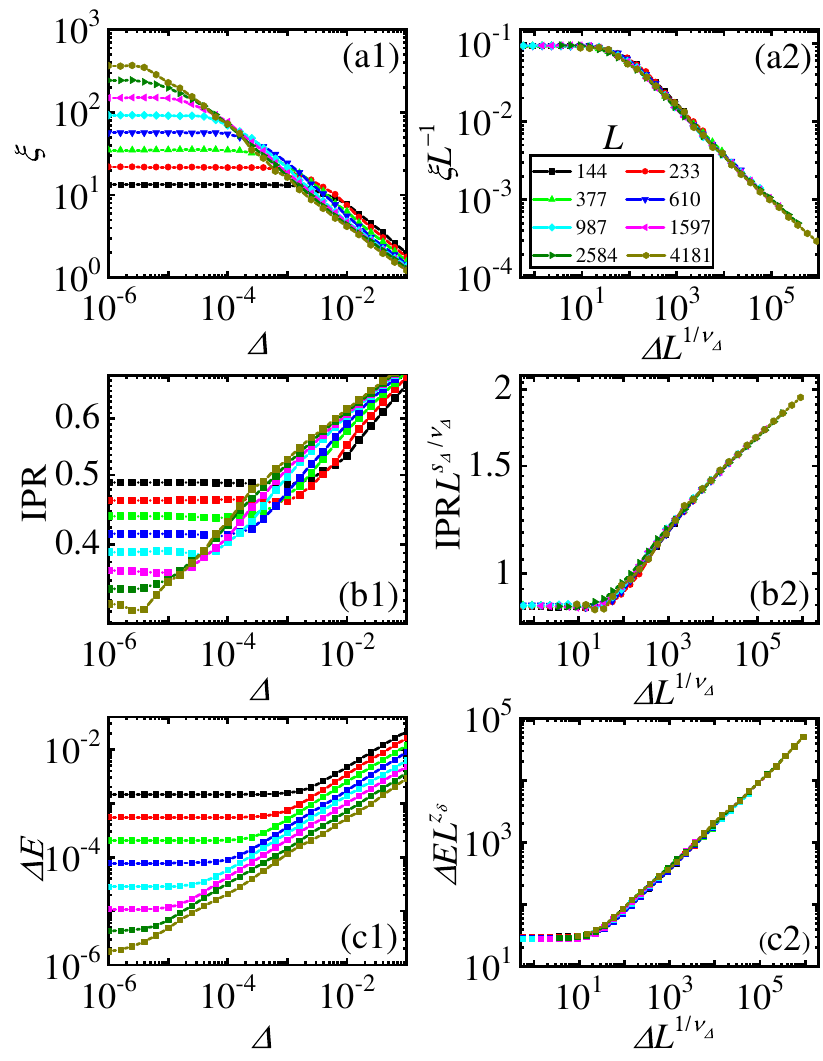}
  \vskip-3mm
  \caption{Scaling properties in the state with the lowest real part of the eigenenergy for fixed $\delta L^{1/\nu_\delta}=-0.5$ under OBC.
  The curves of $\xi$ versus $\Delta$ before (a1) and after (a2) rescaled for different $L'$s.
  The curves of $\rm IPR$ versus $\Delta$ before (b1) and after (b2) rescaled for different $L'$s.
  The curves of $\Delta E$ versus $\Delta$ before (c1) and after (c2) rescaled for different $L'$s.
   Here, $g=0.5$, and the result is averaged for $1000$ samples. The double-logarithmic scales are used.}
  \label{dLn0.5}
\end{figure}
\section*{APPENDIX A: NUMERICAL VALIDATION FOR STATIC SCALING EQUATIONS}
In this Appendix, we carry out a numerical validation of Eqs.~(\ref{Eq:xifull}) to (\ref{Eq:DEfull}) under OBC  with $\delta L^{1/\nu_\delta}$ held constant at -0.5.
Fig.~\ref{dLn0.5} illustrates the scaling behavior of $\xi$, $\rm IPR$ and $\Delta E$ as functions of $\Delta$.
Upon rescaling these quantities in accordance with Eqs.~(\ref{Eq:xifull}) to (\ref{Eq:DEfull}), the resultant curves exhibit a remarkable alignment. This alignment serves as compelling evidence for the validity of these equations.

\section*{APPENDIX B: NUMERICAL VALIDATION FOR KZS EQUATIONS}
\begin{figure}[htbp]
\centering
  \includegraphics[width=1\linewidth,clip]{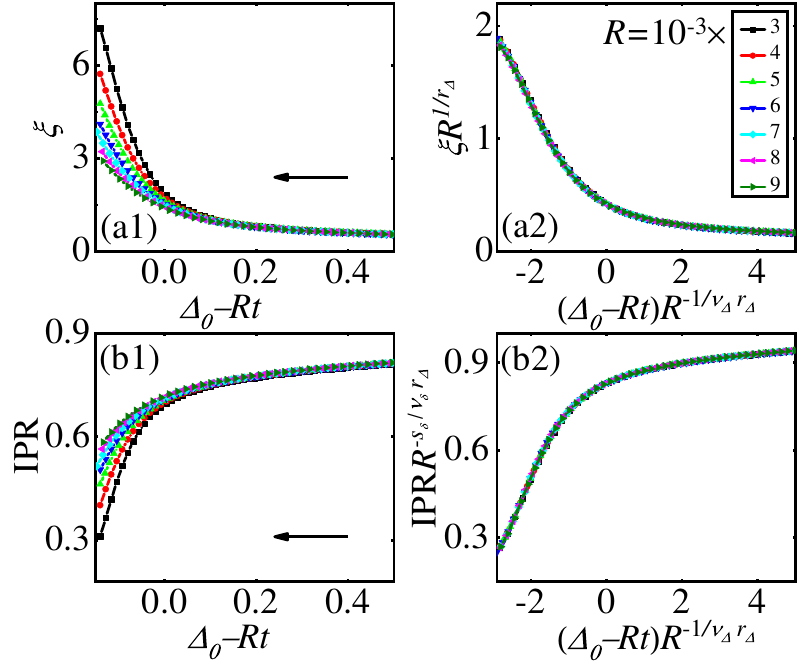}
  \vskip-3mm
  \caption{KZS of driven dynamics in the non-Hermitian DAA model with fixed $\delta R^{-1/r_\Delta\nu_\delta}=0.5$ under OBC.
  Curves of $\xi$ versus $\Delta$ before (a1) and after (a2) rescaling for different driving rates $R$.
  Curves of $\rm IPR$  versus $\Delta$ before (b1) and after (b2) rescaling for different $R$.
  The lattice size is $L=610$, $\phi=0$, $g=0.5$ and one sample of $w_j$ is used.
  $\Delta_0=1.0$.}
  \label{kzsdeltaR05OBC}
\end{figure}

\begin{figure}[htbp]
\centering
  \includegraphics[width=1\linewidth,clip]{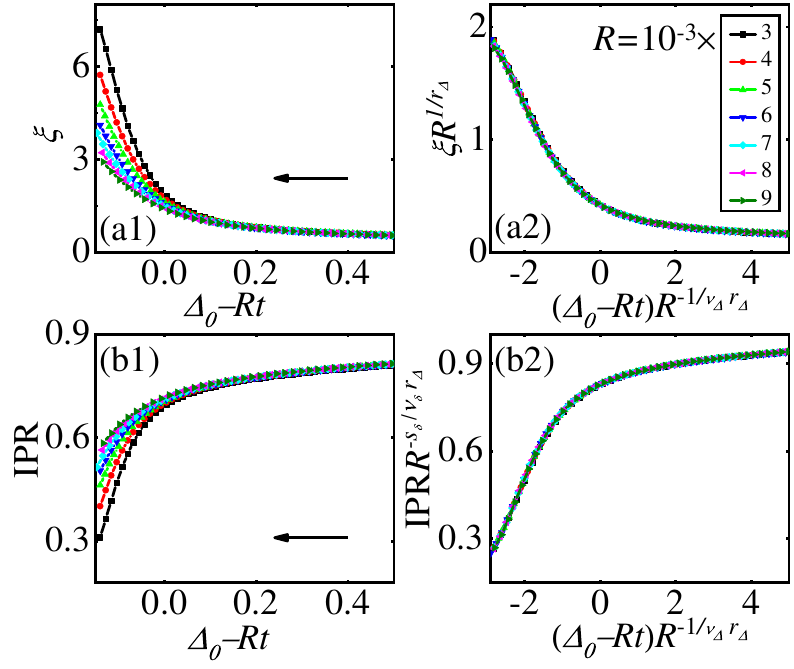}
  \vskip-3mm
  \caption{KZS of driven dynamics in the non-Hermitian DAA model with fixed $\delta R^{-1/r_\Delta\nu_\delta}=0.5$ under PBC.
  Curves of $\xi$ versus $\Delta$ before (a1) and after (a2) rescaling for different driving rates $R$.
  Curves of $\rm IPR$  versus $\Delta$ before (b1) and after (b2) rescaling for different $R$.
  The lattice size is $L=610$, $\phi=0$, $g=0.5$ and one sample of $w_j$ is used.
  $\Delta_0=1.0$.}
  \label{kzsdeltaR05PBC}
\end{figure}

\begin{figure}[htbp]
\centering
  \includegraphics[width=1\linewidth,clip]{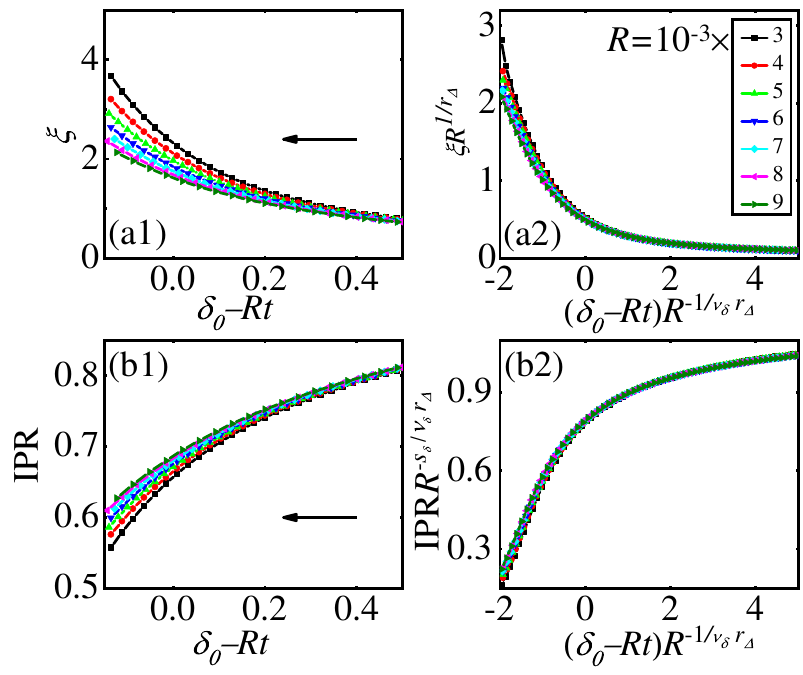}
  \vskip-3mm
  \caption{KZS of driven dynamics in the non-Hermitian DAA model with fixed $\Delta R^{-1/r_\Delta\nu_\Delta}=0.5$ under OBC.
  Curves of $\xi$ versus $\delta$ before (a1) and after (a2) rescaling for different driving rates $R$.
  Curves of $\rm IPR$  versus $\delta$ before (b1) and after (b2) rescaling for different $R$.
  The lattice size is $L=610$, $\phi=0$, $g=0.5$ and one sample of $w_j$ is used.
  $\delta_0=1.0$.}
  \label{kzsDeltaR05OBC}
\end{figure}

\begin{figure}[htbp]
\centering
  \includegraphics[width=1\linewidth,clip]{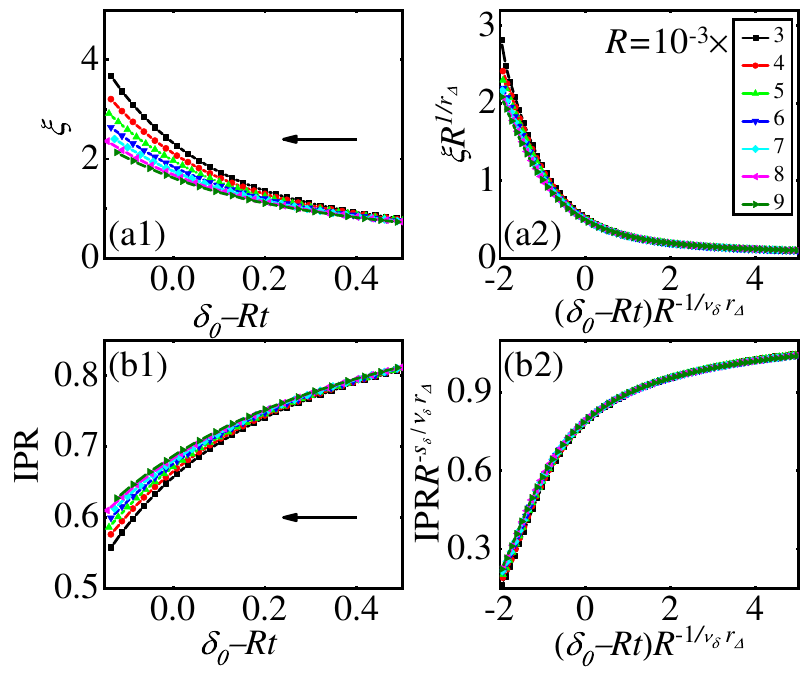}
  \vskip-3mm
  \caption{KZS of driven dynamics in the non-Hermitian DAA model with fixed $\Delta R^{-1/r_\Delta\nu_\Delta}=0.5$ under PBC.
  Curves of $\xi$ versus $\delta$ before (a1) and after (a2) rescaling for different driving rates $R$.
  Curves of $\rm IPR$  versus $\delta$ before (b1) and after (b2) rescaling for different $R$.
  The lattice size is $L=610$, $\phi=0$, $g=0.5$ and one sample of $w_j$ is used.
  $\delta_0=1.0$.}
  \label{kzsDeltaR05PBC}
\end{figure}

In this Appendix, we numerically verify the KZS scaling equations (Eq.~(\ref{Eq:xifullKZS}) to Eq.~(\ref{Eq:IPRfullKZS})) by fixing $\delta R^{-1/r_\Delta\nu_\delta}=0.5$ or $\Delta R^{-1/r_\Delta\nu_\Delta}=0.5$ under both OBC and PBC respectively.
We compute the time evolutions of $\xi$ and $\rm IPR$ for various driving rates $R$, with the initial state with the lowest real part of eigenenergy, and observe that after rescaling according to the KZS equations, the curves for different R converge near the critical region, confirming the validity of these equations as demonstrated in Fig.~\ref{kzsdeltaR05OBC}, Fig.~\ref{kzsdeltaR05PBC}, Fig.~\ref{kzsDeltaR05OBC}, and Fig.~\ref{kzsDeltaR05PBC}.

\vskip-2mm
\vskip-2mm

\section*{}

\end{document}